\documentclass[twocolumn,preprintnumbers,amsmath,amssymb]{revtex4}
\usepackage{graphicx}
\usepackage{dcolumn}
\usepackage{bm}

\begin{document}
\title{
 Graphene Tunneling 
 Transit-Time Terahertz Oscillator Based on Electrically Induced p-i-n 
Junction  
}
\author{V.~Ryzhii\footnote{Electronic mail: v-ryzhii@u-aizu.ac.jp} 
and M.~Ryzhii
}
\affiliation{
Computational Nanoelectronics Laboratory, University of Aizu, 
Aizu-Wakamatsu 965-8580, Japan\\
Japan Science and Technology Agency, CREST, Tokyo 107-0075, Japan
}
\author{V.~Mitin}
\affiliation{Department of Electrical Engineering, University at Buffalo,
State University of New York,
NY 14260, USA}
\author{M.~S.~Shur}
\affiliation{Department of Electrical, Electronics, and System
Engineering, Rensselaer Politechnic Institute,
Troy, NY 12180, USA}
\date{\today}    
\begin{abstract}
We propose and analize a graphene tunneling transit time 
device based on a  
heterostructure with a lateral  p-i-n junction  electrically induced
in the graphene layer by the applied  gate voltages of different polarity.
The depleted i-section of the graphene layer (between the gates) serves as both
the tunneling
injector and the transit region.
Using the developed device model, we demonstrate that the ballistic transit 
of electrons and holes generated due to interband tunneling in the i-section
results in the negative ac conductance
in the terahertz frequency range, so that the device  can serve as a terahertz oscillator.
\end{abstract}

\maketitle

{\it Graphene}   
is considered as a promising candidate for different future electronic and 
optoelectronic
devices.
Its most distinctive features beneficial for device applications
include high electron and hole mobilities
in a wide range of temperatures, 
the possibility of bandgap engineering (creation of the graphene-based structures
with the energy gap from zero to fairly large values),
formation of electrically induced lateral p-n junctions 
(see, for instance,~\cite{1,2,3,4,5}).

The operation of graphene field-effect transistors (G-FETs)
 is accompanied by the formation
of the lateral  n-p-n  (or p-n-p) junction under 
the controlling (top)
gate and the pertinent energy barrier~\cite{6,7,8,9}.
The tunneling across such a n-p-n junction prevents the achievement
of a low off-state current~\cite{9}. This limits possible realization
 of G-FETs  in large scale digital
electronic circuits and forces to consider the graphene structures in which
the energy gap is reinstated (graphene nanoribbons 
and graphene bilayers with 
the energy gap open by 
the transverse electric field)~\cite{10,11,12,13,14,15,16}.

Unique  properties of graphene already produced not only using peeling
technology but also epitaxial methods as well as graphene nanoribbons
and bilayers, particularly, experimental evidences of
the possibility of  
 ballistic electron and hole transport in
samples  with several micrometer sizes even at room temperatures
(see, for instance, Refs.~\cite{17,18})
stimulate inventions of different graphene-based devices which
could not be realized the past using the customary materials.


In this paper, we propose a transit-time oscillator 
which can operate in the terahertz (THz) range of frequencies
and substantiate the 
operational principle of the device. 
The operation of the device in question is associated with
the tunneling electron injection in an electrically induced reverse biased
  lateral 
p-i-n junction
and the electron and hole 
transit-time effects in its depleted section. In the following, this device is
  referred to 
as
the graphene tunneling transit-time (G-TUNNETT) terahertz oscillator.
The tunneling generation through the zero energy gap
and propagation of electrons and holes with their directed velocity $v_x$
close to the characteristic velocity $v_W \simeq 10^8$~cm/s of
 the  graphene energy spectrum
($v_x \sim v_W$),  can provides significant advantages of 
G-TUNNETTs in comparison
with the existing and discussed TUNNETTs based 
on the conventional semiconductor
materials
(see, for instance, Refs.~\cite{19,20} and the references therein).
Using the developed device model, we calculate
the high-frequency characteristics. 
The device under consideration comprises a graphene layer with the
source and drain contacts and the gates. 
The lateral p-i-n junction can be formed  by two top gates
biased by the dc voltages $V_p$ and $V_n$ of different polarities
($V_p < 0$, $V_n > 0$). 
 The G-TUNNETT  structure under consideration
and its band diagram (at the applied source drain-voltage
$V$, so that the lateral  p-i-n junction is reverse biased) are schematically
shown in Fig.~1a and Fig.~1b,
respectively.
It is assumed that apart from a dc component $V_0 > 0$
which corresponds to a recerse bias, the net source-drain voltage 
$V$ comprises also an ac component $\delta V\exp(-i\omega\,t)$:
$V = V_O + \delta V\exp(-i\omega\,t)$, where $\delta\,V$ and $\omega$ are
the signal amplitude and frequency, respectively.

Thus, the graphene layer is partitioned into three sections:
p- and n- sections adjacent to the source and drain contacts, respectively,
and depleted section in the center (i-section). 
The depleted section of the graphene layer plays the dual role:
(1) a strong electric field in the i-section provides the tunneling generation
of electrons and holes and (2) this section serves as the transit region
where the generated electrons and holes propagating ballistically
across this region induce
the current in the  p- and n- sections (as well as, possibly in the contacts)
and, hence, the terminal current 
in the
source-drain circuit.
Considering for definiteness  the "symmetric" device structure 
with $V_n = - V_p = V_t$) corresponding to Fig.~1,
we shall calculate its source-drain  ac conductance 
$\sigma_{\omega}^{sd}$ (dynamic conductance)
as a function
of the structural parameters and the signal frequency and demonstrate
that $\sigma_{\omega}^{sd} < 0$ in certain (THz) frequency ranges. 

\begin{figure}[t]
\vspace*{-0.4cm}
\begin{center}
\includegraphics[width=7.5cm]{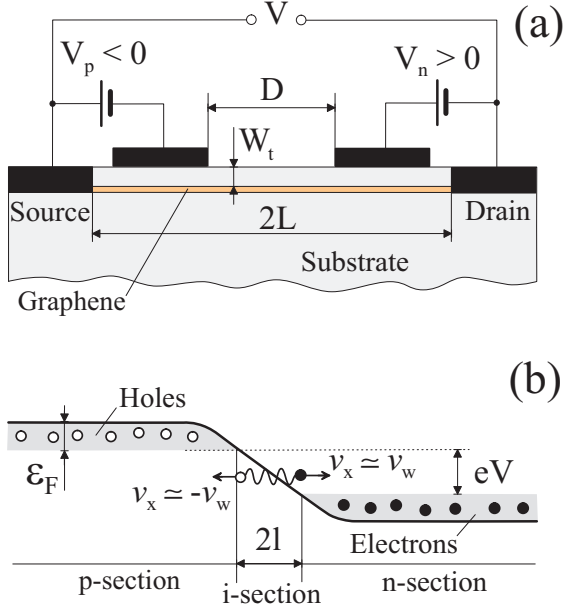}
\caption{Schematic view  of a G-TUNNETT  structure (a)
and its band diagram (b).
}
\end{center}
\end{figure}

Using the value of the  interband tunneling probability~\cite{5,21},
and roughly estimating the electric field in the i-section as
${\cal E} \simeq V/2l$, where $2l$ is the length of the depleted
i-section ($2l \simeq D$ if  the spacing between 
the top gates $D \gg W_t$, were $W_t$ is the thickness of 
the layers separating
the graphene layer and the gates;  more strict and 
detailed calculations
 can be found 
in Refs.~\cite{21,22}), for 
the rate of the tunneling generation of electron hole pair (per unit length in the transverse direction) in the i-section
 in the situation under consideration one can obtain
\begin{equation}\label{eq1}
G = \frac{g_0}{8\pi}\sqrt{\frac{{\cal E}}{e\hbar\,v_W}}V
\simeq  \frac{g_0}{8\pi\sqrt{2e\hbar\,v_Wl}}V^{3/2}.
\end{equation}
Here $g_0 =2e^2/\pi\hbar$, $e$ is the electron charge, 
and $\hbar$
is the reduced Planck constant.
This corresponds to the source-drain dc current
\begin{equation}\label{eq2}
 J_0 = 2eG_0 = \frac{g_0}{4\pi}\sqrt{\frac{e}{2\hbar\,v_Wl}}V_0^{3/2}.
\end{equation}
The ac component of the generation rate is given by
\begin{equation}\label{eq3}
\delta G_{\omega} \simeq  \frac{3g_0}{16\pi\sqrt{2e\hbar\,v_Wl}}V_0^{1/2}
\delta V_{\omega} = \frac{3}{2} \frac{J_0}{e}\frac{\delta V_{\omega}}{V_0}.
\end{equation}

The tunneling probability $w$ is a fairly sharp function of the angle
between the direction of the electron (hole) motion and
the x-direction (from the source to the drain):
$w(\theta) \simeq \exp(- \alpha\sin^2 \theta)$,
where $\alpha \propto D $ is rather large~\cite{5}. 
Considering this, one can disregard some
spread in the x-component of the velocity ($\Delta\,v_x/v_W \simeq 1/2\alpha$)
of the injected electrons
and assume that all the generated electrons and holes propagate
in the x-direction with the velocity $v_x \simeq v_W$.

In the case of ballistic electron and hole transport
in the i-section when the  generated   electrons and holes
do not change the directions of their propagation,
the continuity equations  governing
the ac components of the electrons and holes densities $\delta\Sigma_{\omega}^-$
and $\delta\Sigma_{\omega}^+$
can be presented as:
\begin{equation}\label{eq4}
-i\omega\delta\Sigma_{\omega}^{\mp} \pm v_W\frac{d\,\delta\Sigma_{\omega}^{\mp}}{d\,x}
= \frac{\delta G_{\omega}}{2l},
\end{equation}
The boundary conditions are as follows: $\delta\Sigma_{\omega}^{\mp}|_{x = \mp l} = 0$.
Solving Eq.~(4) with these boundary conditions, we arrive at
\begin{equation}\label{eq5}
\delta\Sigma_{\omega}^{\mp}(x) = \frac{\delta G_{\omega}^T}{2l}.
\frac{\exp[i\omega(l \pm x)/v_W] - 1}{i\omega}
\end{equation}

Considering this, for
the net  ac current  $\delta\,J_{\omega}(x) = ev_W[\delta\Sigma_{\omega}^{-}(x)
+ \delta\Sigma_{\omega}^{+}(x)]$, created by both the generated
electrons and holes, one can obtain
\begin{equation}\label{eq6}
\delta\,J_{\omega}(x) = 
e\delta G_{\omega}^T\biggl[\frac{e^{\displaystyle i\omega\tau}
\cos(\omega\tau\,x/l) - 1}{i\omega\tau}\biggr].
\end{equation}
Here, we have introduced the characteristic transit time $\tau = l/v_W$
Since the real  time of the electron and hole transit across the i-section
varies from $0$ to $2l/v_W$, $\tau$ is actually the mean transit time.
If the signal frequency
$\omega$ is smaller than the plasma oscillations in the p- and n-sections
$\Omega$, ($\omega < \Omega$),
these sections can be considered as just highly conducting electrodes,
so that the terminal ac current is mainly induced in them.
In this case,
the ac component of the source-drain terminal current $\delta J_{\omega}^{sd}$,
i.e., the ac current induced by the propagating electrons in the external
circuit connecting the source and the drain 
can, according to the Ramo-Shockley theorem~\cite{23,24},  be presented as
$\delta J_{\omega}^{sd} = \int_{-l}^{l}d\,x\,g(x) \delta J_{\omega}(x) - 
i\omega\,C\delta V_{\omega}$,
where 
$g(x)$ is the form-factor determined by the shape of highly conducting
regions (the p-section and the drain contact) and $C$ is the i-section geometrical
capacitance.
The factor $g(x)$ is associated with different contributions
to the induced current 
of the electrons and holes at different distances from the highly conducting
regions. 
For the bulky conducting regions $g(x) \simeq 1/2l$,
whereas for the blade-like conducting regions~\cite{25}, 
$g(x) = 1/\pi\sqrt{l^2 - x^2}$. For the i-section capacitance one can use
the following formula:~\cite{26} $C = (\ae/4\pi^2)\ln(4L/l)$, where
$\ae$ is the dielectric constant.
Considering Eqs.~(3), we arrive at the following formula for the ac conductance
$\sigma_{\omega}^{sd} = \delta J_{\omega}^{sd}/\delta V_{\omega}$:
$$
\sigma_{\omega}^{sd} = \frac{3\sigma_{0}}{2\pi}\int_{-1}^{1}
\frac{d\xi }{\sqrt{1 - \xi^2}}\biggl[\frac{e^{\displaystyle i\omega\tau}
\cos(\omega\tau\xi) - 1}{i\omega\tau}\biggr] - i\omega\,C
$$
\begin{equation}\label{eq7}
 = \frac{3\sigma_{0}}{2}
\biggl[\frac{e^{\displaystyle i\omega\tau}
{\cal J}_0(\omega\tau) - 1}{i\omega\tau}\biggr] - i\omega\,C,
\end{equation}
where $ \sigma_{0}  = J_0/V_0$
is the dc conductance and  ${\cal J}_0(\xi)$ is the Bessel function.

The real and imaginary parts of the ac conductance are given by
\begin{equation}\label{eq8}
{\rm Re}\sigma_{\omega}^{sd} = \frac{3\sigma_{0}}{2}
\frac{\sin(\omega\tau)}{\omega\tau}\, {\cal J}_0(\omega\tau),
\end{equation}
\begin{equation}\label{eq9}
 {\rm Im} \sigma_{\omega}^{sd}= \frac{3\sigma_{0}}{2}\biggl[
\frac{1 - \cos(\omega\tau){\cal J}_0(\omega\tau)}{\omega\tau} - c\, \omega\tau\biggr],
\end{equation}
where $c = 2C/3\sigma_0\tau$.
At relatively low signal frequencies ($\omega\tau \ll 1$),
Eqs.~(7) and (8) yield Re~$\sigma_{\omega}^{sd} \simeq (3\sigma_{0}/2)
(1 - 7\omega^2\tau^2/12)$ and 
Im~$\sigma_{\omega}^{sd} \simeq  (3\sigma_{0})(3/4 - c)\omega\tau$, respectively.
Thus, 
$\sigma_{0}^{sd} \simeq 3\sigma_{0}/2$, 
i.e., is equal
to the differential dc conductance 
$\sigma^{sd}_{diff} = (\partial J_0/\partial V_0)|_{V_t = const}$.
Assuming that $2l = 0.5~\mu$m, $2L = 2~\mu$m, $\ae = 3.8$, and $V_0 = 0.1 - 1$~V,
we obtain $\sigma_0 \simeq (2.09 - 6.64)\times 10^{12}$~s$^{-1}$ and $c \simeq 0.34$.
\begin{figure}[t]
\vspace*{-0.4cm}
\begin{center}
\includegraphics[width=7.5cm]{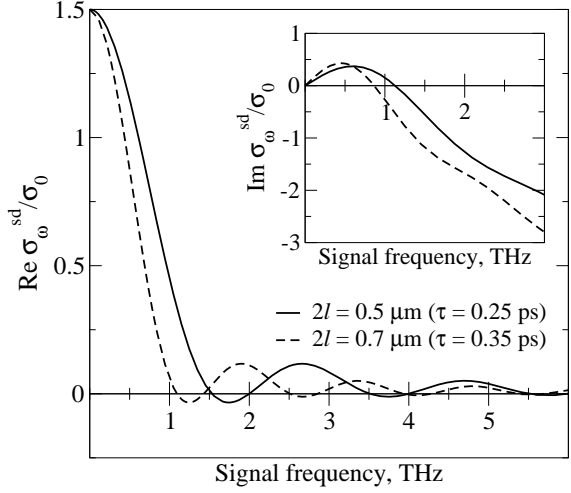}
\caption{Real part and imaginary part (inset) of the ac conductance
$\sigma_{\omega}^{sd}$ as functions 
of signal frequency
$f = \omega/2\pi$ for different lengths of i-section (different transit times).
}
\end{center}
\end{figure}
The real part of the ac conductance Re $\sigma_{\omega}^{sd}$
turns zero at the following signal frequencies
$f = \omega/2\pi$:
$f^{-}_1 \simeq 0.38/\tau$, $f^{+}_1 = 0.5/\tau$, 
$f^{-}_2 \simeq 0.88/\tau$,  $f^{+}_2 = 1/\tau$,...
In the frequency ranges $0.38/\tau < f < 0.5/\tau$ and $0.88/\tau < f < 1/\tau$,
the real part of the ac conductance is negative.
In particular, at $2l = 0.7~\mu$m ($\tau = 0.35$~ps),
$f^{-}_1 \simeq 1.08$~THz and $f^{+}_1 = 1.43$~THz.
The real part of the ac conductance ${\rm Re} \sigma_{\omega}^{sd}$
reaches minima at certain frequencies $f_1, f_2,...$, 
which fall into the intervals $f^{-}_1 < f_1 < f^{+}_1$,
$f^{-}_2 < f_2 < f^{+}_2$,...
The quantity ${\rm Re}\sigma_{\omega}^{sd}$ increases with increasing
source-drain voltage and, consequently,
dc current. A decrease in the i-section length at fixed source-drain voltage
gives rise to  an increase  both in    $|{\rm Re} \sigma_{\omega}^{sd}|$ and
in the frequencies  $f_1, f_2,...$, 
where ${\rm Re} \sigma_{\omega}^{sd}$ exhibits manima. 
 
Figure~2 shows the real and imaginary parts of the ac conductance 
$\sigma_{\omega}^{sd}$ as functions
of the signal frequency $f = \omega/2\pi$ calculated for
$2l = 0.5~\mu$m and  $2l = 0.7~\mu$m. 
The real part exhibits a pronounced oscillatory behavior
with the frequency ranges where it has different signs.
The frequency dependence of the imaginary part corresponds to
the domination of  the inductive component 
at low frequencies (associated with the  contribution of electrons and holes)
and the capacitive component at elevated frequencies (due to the contribution of the geometrical capacitance). The relative value of the real part of the
ac conductance at the first minima at $2l = 0.5~\mu$m
is fairly moderate:   ${\rm Re}\sigma_{\omega = 2\pi\,f_1}^{sd}/\sigma_0 \simeq  - 0.034$.
However, it is markedly larger than that in some new concept 
THz devices~\cite{27}.
Considering the above estimate of $\sigma_0$ at $V_0 = 0.1 - 1$~V, one can obtain  
 ${\rm Re}\sigma_{\omega = 2\pi\,f_1}^{sd}\simeq - (0.71 - 2.26)\times 10^{11}$~s$^{-1}$.
As follows from the above estimates and Fig.~2, the negativity of the real
part of the ac conductance at the  frequencies $f > 1$~THz
can be achieved 
in the  G-TUNNETT structures with the i-section length  only moderately 
smaller than one  micrometer. 
We assumed above that the generated electrons and holes in the i-section
propagate ballistically. Taking into account that the electron-electron 
and electron-hole
scattering in this  (depleted) section  should be weak,
the mean-free paths in graphene can be of several micrometers even 
at room temperatures~\cite{17,18}. 
Our estimates show that the effect of the  electron  and hole space charge
 on the 
potential distribution 
in the i-section is
weak 
 and the pertinent plasma phenomena are insignificant. 
\begin{figure}[t]
\vspace*{-0.4cm}
\begin{center}
\includegraphics[width=7.5cm]{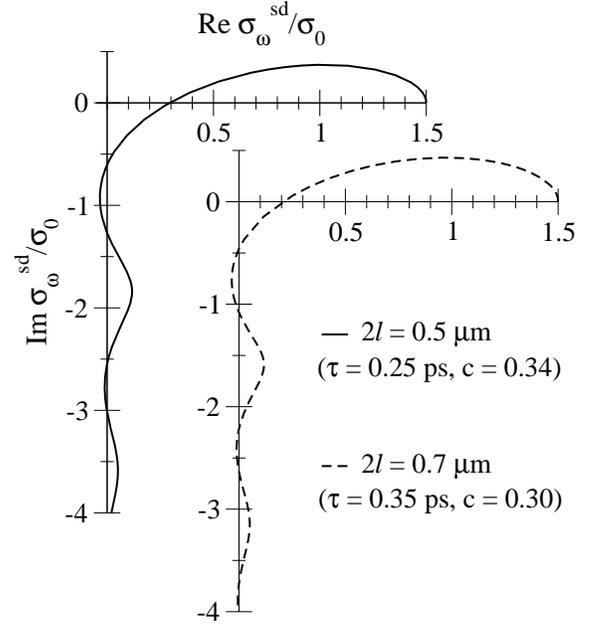}
\caption{Real part vs  imaginary part 
dependence with $\omega\tau$ as a parameter
 for different length of i-section (different transit times and normalized capacitances).
}
\end{center}
\end{figure}

The fundamental plasma frequency in the gated p- and n-sections,
in which the spectrum of the plasma waves is a sound-like 
$\Omega = \pi\,s/2(L - l)$, where $s$ is
the characteristic velocity, which at sufficiently high gate voltages
when the electron (hole) Fermi energy in the n-section  (p-section)
$\varepsilon_{F} = (\hbar\,v_W/2\sqrt{2})\sqrt{\ae\,V_t/eW_t} \gg k_BT$,
is given by 
$s\simeq \sqrt{2e^2W_t\varepsilon_F/\ae\hbar^2} > v_W$~\cite{28}, 
$2L$ is the spacing between the source and drain contacts,
 $T$ is the temperature, and $k_B$ is the Boltzmann
constant. Since the the plasma frequency in graphene can be rather large
(owing to a large $s$),
the condition  $\omega < \Omega$ assumed above can be easily fulfilled
for the signal frequencies in the THz range 
(as well as the condition $\omega \simeq \Omega$) by a proper choice of the device
structure length $2L$ and the gate voltage $V_t$.
Indeed, setting, for example,  $s = (3 - 6)\times10^8$~cm/s, $2l = 0.5~\mu$m,
and  $2L = 1.5 - 2~\mu$m,
one can obtain $\Omega/2\pi = 1 - 3$~THz.

If $f_1^{-}, f_1^{+} < \Omega/2\pi$ (or  $f_2^{-}, f_2^{+} < \Omega/2\pi$),
a G-TUNNETT coupled with the proper resonant cavity can serve as a 
THz oscillator.
However, when  $f_1^{-} < \Omega/2\pi < f_1^{+} $ 
(or  $f_2^{-} < \Omega/2\pi < f_2^{+}$), i.e., when the frequency of the plasma
oscillations falls into the range where Re~$\sigma_{\omega}^{sd} < 0$,
the hole and electron systems in the gated p- and n-sections, respectively,
can play the role of the resonators. In this situation, the self-excitation
of the plasma oscillations (plasma instability) can be possible
if the quality factor of these oscillations  $Q
\simeq \Omega/\nu$, where $\nu$ is the frequency
of hole and electron  collisions with defects
and acoustic phonons in the p- and n-sections, 
is sufficiently large~\cite{26}
(see also Refs.~\cite{29,30}). It should be noted that
the electron-electron and hole-hole scattering processes do not affect $Q$ except
as via relatively weak  effects of electron and hole viscosity~\cite{31}.
As a result, in sufficiently perfect graphene layers, $\nu$ can be small not
only in the i-section but in the p- and n-sections as well.  
Thus, a G-TUNNETT can work as a THz  source in 
the regime of plasma instability 
as well. In such a case, a  G-TUNNETT should be supplied by an antenna,
although the gates can also play the role of the latter.

In conclusion, we have proposed  a G-TUNNETT and calculated its ac conductance
as a function of the signal frequency and the structural parameter
using the developed device model. We have demonstrated that 
 the ac conductance
exhibits the frequency ranges where it is  negative.
Due to high directed energy independent
velocities of the electrons and holes
generated owing to the interband tunneling,
these frequency regions correspond to the THz range at relatively large length of the
i-region. A G-TUNNETT can work as an  active element of THz oscillators
with a complemetary resonant cavity or can immediately emit THz radiation
(in the plasma instability mode).

The authors are grateful to Professor T.~Otsuji for stimulating discussions.
The work was supported by  the Japan Science and 
Technology Agency, CREST and
by Grant-in-Aid for Scientific Research (S) from the
Japan Society for Promotion of Science.
The work  was also partially 
supported by the Airforce 
Office of Scientific Research, U.S.A.


\end{document}